\documentclass[showpacs,twocolumn,prl,aps, superscriptaddress]{revtex4}
\usepackage{amssymb}
\usepackage{amsmath}
\usepackage{graphicx}

\input{tcilatex}

\begin{document}

\title{TOPOLOGICAL INSULATOR AND THE DIRAC EQUATION}
\date{\today }
\author{Shun-Qing Shen, Wen-Yu Shan and Hai-Zhou Lu \\
Department of Physics, The University of Hong Kong, Pokfulam Road, Hong Kong}
\begin{abstract}
We present a general description of topological insulators from the point of
view of Dirac equations. The Z$_{2}$ index for the Dirac equation is always
zero, and thus the Dirac equation is topologically trivial. After the
quadratic $B$ term in momentum is introduced to correct the mass term $m$ or
the band gap of the Dirac equation, the Z$_{2}$ index is modified as 1 for $%
mB>0$ and 0 for $mB<0$. For a fixed $B$ there exists a topological quantum
phase transition from a topologically trivial system to a non-trivial one
system when the sign of mass $m$ changes. A series of solutions near the
boundary in the modified Dirac equation are obtained, which is
characteristic of topological insulator. From the solutions of the bound
states and the Z$_{2}$ index we establish a relation between the Dirac
equation and topological insulators.
\end{abstract}

\maketitle


\section{Introduction}

Translational invariance in crystal lattices and the Bloch theorem for the
wave function of electrons in solid make it possible for us to know the band
structures of solid and why a solid is a metal, an insulator or
semiconductor. Recent years it is found that a new class of materials
possess a feature that its bulk is insulating while its surface or edge is
metallic. This metallic behavior is quite robust against impurities or
interaction, and is protected by the intrinsic symmetry of the band
structures of electrons. The materials with this new feature is called
topological insulator.\cite{Moore-10Nature,Qi-10PhysTo,Hasan-10RMP}

In 1979 the one-parameter scaling theory predicted that all electrons in
systems of two or lower dimension should be localized for a weak disorder.%
\cite{Abrahams-79prl} This theory shaped the research of lower dimensional
systems with disorders or interaction. Almost at the same time, von Klitzing
et al discovered experimentally integer quantum Hall effect (IQHE) in
two-dimensional (2D) electron gas in semiconductor hetero-junction in a
strong magnetic field, in which longitudinal conductance becomes zero while
the quantum plateau of the Hall conductance appears at $\nu e^{2}/h$ ($\nu $
is an integer).\cite{Klitzing-80prl} Two years later Tsui et al observed the
fractional quantum Hall effect (FQHE) in a sample with higher mobility.\cite%
{Tsui-82prl} In the theory of edge states for IQHE electrons form discrete
Landau levels in a strong magnetic field. Electrons in the bulk has a
vanishing group velocity, and are easily localized by impurities or disorder
while the electrons near the boundary are skipping along the boundary to
form a conducting channel.\cite{Laughlin-81prb,Halperin-82prb} Thus in IQHE
all bulk electrons are localized to be insulating while the edge electrons
form several conducting channels according to the electron density which is
robust against the impurities. This feature indicates that IQHE is a new
state of quantum matter, i.e., one of topological insulators. In FQHE it is
the electron-electron interaction that makes electrons incompressible and to
form stable metallic edge states.\cite{Laughlin-83prl,Jain-89prl} The
quasiparticles in FQHE have fractionalized charges, and obeys new quantum
statistics. In 1988 Haldane proposed that IQHE could be realized in a
lattice system of spinless fermions in a period magnetic flux.\cite%
{Haldane-88prl} Though the total magnetic flux is zero, electrons are driven
to form an conducting edge channel by the magnetic flux. Since there is no
pure magnetic field the quantum Hall conductance originates from the band
structure of fermions in the lattice instead of the discrete Landau levels.

In 2005 Kane and Mele generalized the Haldane's model to a lattice of spin
1/2 electrons.\cite{Kane-05prl} The strong spin-orbit coupling, an effect of
relativistic quantum mechanics for electrons in atoms, is introduced to
replace the periodic magnetic flux in Haldane's model. This interaction
looks like an spin-dependent magnetic field to employ on electron spins.
Different electron spins experience opposite spin-orbit force, i.e., spin
transverse force.\cite{Shen-05prl} As a result, a bilayer Haldane model may
be realized in a spin-1/2 electron system with spin-orbit coupling, which
exhibits quantum spin Hall effect (QSHE). In the case there exist
spin-dependent edge states around the boundary of the system: electrons with
different spins move in opposite directions, and form a pair of helical edge
states. In this system the time reversal symmetry is preserved, and the edge
states are robust against impurities or disorders because the electron
backscattering in the two edge channel is prohibited due to the symmetry.
Bernevig, Hughes and Zhang predicted that the QSHE effect can be realized in
the HgTe/CdTe sandwiched quantum well.\cite{Bernevig-06science} HgTe is an
inverted-band material, and CdTe is a normal band one. Tuning the thickness
of HgTe layer may lead to the band inversion in the quantum well, which
exhibits a topological phase transition.\cite{Shen-04prb,Zhou-07epl} This
prediction was confirmed experimentally by Konig et al soon after the
prediction.\cite{Konig-07science} The stability of the QSHE was studied by
several groups.\cite{Sheng-06prl,Xu-06prb,Wu-06prl,Onoda-07prl} Li et al
discovered that the disorder may even generate QSHE, and proposed a possible
realization of topological Anderson insulator, in which all bulk electrons
are localized by impurities meanwhile an conducting helical edge channels
appear.\cite{Li-09prl} This phase was studied numerically and analytically.%
\cite{Jiang-09prb,Groth-09prl} Strong Coulomb interactions may also generate
QSHE in Mott insulators.\cite{Mong-10prb,Pesin-10np}

The generalization of QSHE to three dimension is topologically non-trivial.%
\cite{Fu-07prl,Moore-07prb,Murakami-07NJP,Roy-09prb} Kane and Mele proposed
a Z$_{2}$ index to classify the materials with time reversal invariance into
a strong and weak topological insulator.\cite{Kane-05prlb} For a strong
topological insulator, there exists an odd number of surface states crossing
the Fermi surface of the system. The backscattering of electrons in the
surface states are prohibited because of the symmetry. Bi$_{1-x}$Sb$_{x}$
was predicted to be 3D topological insulator by Fu and Kane\cite{Fu-07prb}
and was verified experimentally.\cite{Hsieh-08nature} Zhang et al \cite%
{Zhang-09NP} and Xia et al\cite{Xia-09NP} pointed out that Bi$_{2}$Te$_{3}$
and Bi$_{2}$Se$_{3}$ are topological insulator with a single Dirac cone of
the surface states. ARPES data showed clearly the existence of single Dirac
cone in Bi$_{2}$Se$_{3}$ \cite{Xia-09NP} and Bi$_{2}$Te$_{3}$ \cite%
{Chen-09science}. Electrons in the surface states possess a quantum spin
texture structure, and electron momenta are coupled strongly with electron
spins. These result in a lot of exotic magnetoelectric properties. Qi et al. 
\cite{Qi-08prb}proposed the unconventional magneto-electric effect for the
surface states, in which electric and magnetic fields are coupled together
and are governed by so-call "axion equation" instead of Maxwell eqautions.
It is regarded as one of the characteristic features of the topological
insulators~\cite{Qi-09Science,Essin-09prl}. Fu and Kane proposed a possible
realization of Majorana fermions as an proximity effect of s-wave
superconductor and surface states of topological insulator.\cite{Fu-08prl}
The Majorana fermions are topologically protected from local sources of
decoherence, and will be of potential application in universal quantum
computor.\cite{Freedman-02cmp,Kitaev-06ap} Thus the topological insulators
open a new route to explore novel and exotic quantum particles in condensed
matters.

The Dirac equation is a relativistic quantum mechanical wave function for
elementary spin 1/2 particle.\cite{Dirac-1928,Dirac-book} It enters the
field of topological insulator in two aspects. First of all, topological
insulators possess strong spin-orbit coupling, which is a consequence of the
Dirac equation.\cite{Winkler-book} It makes the spin, momentum and the
Coulomb interaction or external electric fields couple together. As a result
it is possible that the band structures in some materials becomes
topologically non-trivial. Another aspect is that the effective Hamiltonians
to the QSHE and 3D topological insulators have the identical mathematical
structure of the Dirac equation. In these effective models the equations are
used to describe the coupling between electrons the conduction and valence
bands, not the electron and positions in Dirac's theory. The positive and
negative spectra are for the electrons and hole in semiconductors not in the
high energy physics. The conventional Dirac equation is time-reversal
invariant. For a system with time reversal symmetry, the effective
Hamiltonian to describe the electrons near the Fermi level can be derived
from the theory of invariants. As a result of the $k\cdot p$ expansion of
the band structure, some effective continuous models have the same form of
the Dirac equation.

In this paper we start with the Dirac equation to provide a simple but
unified description for a large family of topological insulators. A\ series
of solvable differential equations are presented to demonstrate the
existence of edge and surface states in topological insulators.

\section{Dirac Equation and Solutions of the Bound States}

In 1928, Paul A. M. Dirac wrote down an equation for relativistic quantum
mechanical wave functions, which describes elementary spin-1/2 particles,%
\cite{Dirac-1928,Dirac-book}

\begin{equation}
H=cp\cdot \alpha +mc^{2}\beta
\end{equation}%
where $m$ is the rest mass of particle and $c$ is the speed of light. $%
\alpha _{i}$ and $\beta $ are the Dirac matrices satisfying 
\begin{subequations}
\begin{eqnarray}
\alpha _{i}^{2} &=&\beta ^{2}=1 \\
\alpha _{i}\alpha _{j} &=&-\alpha _{j}\alpha _{i} \\
\alpha _{i}\beta &=&-\beta \alpha _{i}
\end{eqnarray}%
In 2D spatial space, the Dirac matrices have the same forms of the Pauli
matrices $\sigma _{i}$, i.e., $\alpha _{x}=\sigma _{x}$, $\alpha _{y}=\sigma
_{y}$, and $\beta =\sigma _{z}$. In three dimensional spatial space, one
representation of the Dirac matrices in terms of the Pauli matrices $\sigma
_{i}$ ($i=x,y,z$) is 
\end{subequations}
\begin{subequations}
\begin{align}
\alpha _{i}& =\left( 
\begin{array}{cc}
0 & \sigma _{i} \\ 
\sigma _{i} & 0%
\end{array}%
\right) \\
\beta & =\left( 
\begin{array}{cc}
\sigma _{0} & 0 \\ 
0 & -\sigma _{0}%
\end{array}%
\right)
\end{align}%
where $\sigma _{0}$ is a $2\times 2$ identity matrix. From this equation,
the Einstein's relativistic energy-momentum relation will be automatically
the solution of the equation, $E^{2}=m^{2}c^{4}+p^{2}c^{2}.$ This equation
demands the existence of antiparticle, i.e. particle with negative energy or
mass, and predates the discovery of position, the antiparticle of the
electron. It is one of the main achievements of modern theoretical physics.

Under the transformation of $m\rightarrow -m$ it is found that the equation
remains invariant if $\beta \rightarrow -\beta ,$ which satisfies all mutual
anticommutation relations for $\alpha _{i}$ and $\beta $. This reflects the
symmetry between the positive and negative energy particles.

Possible relation between the Dirac equation and the topological insulator
can be seen from a solution of the bound state at the interface between two
regions of positive and negative masses. For simplicity, we first consider a
one-dimensional (1D) example 
\end{subequations}
\begin{equation}
h(x)=-iv\hbar \partial _{x}\sigma _{x}+m(x)v^{2}\sigma _{z}
\end{equation}%
and 
\begin{equation}
m(x)=\left\{ 
\begin{array}{ccc}
-m_{1} & \text{if} & x<0 \\ 
+m_{2} & \text{otherwise} & 
\end{array}%
\right.
\end{equation}%
(and $m_{1}$ and $m_{2}>0$). Except for the extended solutions in the whole
space, there exists a solution of the bound state with zero energy%
\begin{equation}
\Psi (x)=\sqrt{\frac{v}{\hbar }\frac{m_{1}m_{2}}{m_{1}+m_{2}}}\left( 
\begin{array}{c}
i \\ 
1%
\end{array}%
\right) e^{-\left\vert m(x)vx\right\vert /\hbar }.
\end{equation}%
The solution dominantly distributes near the point of $x=0$ and decays
exponentially away from the point of $x=0$. The solution of $m_{1}=m_{2}$
was first obtained by Jackiw and Rebbi, and is a basis for the
fractionalized charge in one-dimensional system.\cite{Jackiw-76prd}. The
solution exists even when $m_{2}\rightarrow +\infty $. In this case, $\Psi
(x)\rightarrow 0$ for $x>0$. However, we have to point out that the wave
function does not vanish at the interface when $m_{2}\rightarrow +\infty $.
If we regard the vacuum as a system with an infinite positive mass, a system
of a negative mass with an open boundary condition forms a bound state near
the boundary. This is the source of some popular pictures for topological
insulator.

In 2D, we consider a system with an interface parallel to the y-axis, with $%
m(x)=m_{1}$\ for $x>0$, and $-m_{2}$ for $x<0$. $k_{y}$ is a good quantum
number. We have two solutions which the wave functions dominantly
distributes around the interface. One solution has the form%
\begin{equation}
\Psi (x,k_{y})=\sqrt{\frac{v}{h}\frac{m_{1}m_{2}}{m_{1}+m_{2}}}\left( 
\begin{array}{c}
i \\ 
0 \\ 
0 \\ 
1%
\end{array}%
\right) e^{-\left\vert m(x)vx\right\vert /\hbar +ik_{y}y}
\end{equation}%
with the dispersion $\epsilon _{k}=v\hbar k_{y}$. Another one has the form%
\begin{equation}
\Psi (x,k_{y})=\sqrt{\frac{v}{h}\frac{m_{1}m_{2}}{m_{1}+m_{2}}}\left( 
\begin{array}{c}
0 \\ 
i \\ 
1 \\ 
0%
\end{array}%
\right) e^{-\left\vert m(x)vx\right\vert /\hbar +ik_{y}y}
\end{equation}%
with the dispersion $\epsilon _{k}=-v\hbar k_{y}$. Both states carry a
current along the interface, but electrons moving in opposite directions.
The currents decays exponentially away from the interface. As the system
does not break the time reversal symmetry, the two states are counterpart
with time reversal symmetry with each other. This is a pair of helical edge
(or bound) states at the interface.

In 3D, we can also find a solution for the surface states, The dispersion
relation for the surface states are $\epsilon _{p}=\pm vp$. It has a
rotational symmetry and forms a Dirac cone

From these solutions we found that the edge states and surface states exist
at the interface of systems with positive and negative masses. However,
since there is a positive-negative mass symmetry in the Dirac equation, we
cannot simply say which one is topologically trivial or non-trivial. Thus
the Dirac equation alone is not enough to describe the topological
insulators.

\section{Modified Dirac equation and Z2 topological invariant}

To explore the topological insulator, we start with a modified Dirac
Hamiltonian by introducing a quadratic correction $-Bp^{2}$ in momentum $%
\mathbf{p}$ to the band gap or rest-energy term,

\begin{equation}
H=v\mathbf{p}\cdot \alpha +\left( mv^{2}-B\mathbf{p}^{2}\right) \beta .
\label{Dirac-eq}
\end{equation}%
where $mv^{2}$ is the band gap of particle and $m$ and $v$ have dimensions
of mass and speed, respectively. the quadratic term breaks the mass symmetry
in the Dirac equation, and makes this equation topologically distinct from
the original one.

The general solutions of the wave functions can be expressed as $\Psi _{\nu
}=u_{v}(p)e^{i(p\cdot r-E_{p,\nu }t)/\hbar }$. The dispersion relations of
four energy bands are $E_{p,\nu (=1,2)}=-E_{p,\nu (=3,4)}=\sqrt{%
v^{2}p^{2}+(mv^{2}-Bp^{2})^{2}}$. The four-component spinors $u_{v}(p)$ can
be expressed as $u_{v}(p)=Su_{\nu }(p=0)$ with%
\begin{equation}
S=\sqrt{\frac{\epsilon _{p}}{2E_{p,1}}}\left( 
\begin{array}{cccc}
1 & 0 & -\frac{p_{z}v}{\epsilon _{p}} & -\frac{p_{-}v}{\epsilon _{p}} \\ 
0 & 1 & -\frac{p_{+}v}{\epsilon _{p}} & \frac{p_{z}v}{\epsilon _{p}} \\ 
\frac{p_{z}v}{\epsilon _{p}} & \frac{p_{-}v}{\epsilon _{p}} & 1 & 0 \\ 
\frac{p_{+}v}{\epsilon _{p}} & -\frac{p_{z}v}{\epsilon _{p}} & 0 & 1%
\end{array}%
\right)
\end{equation}%
where $p_{\pm }=p_{x}\pm ip_{y}$, $\epsilon _{p}=E_{p,1}+\left(
mv^{2}-Bp^{2}\right) $, and $u_{\nu }(0)$ is one of the four eigen states of 
$\beta $.

The topological properties of the modified Dirac equation can be gained from
these solutions of a free particle. The Dirac equation is invariant under
the time-reversal symmetry, and can be classified according to the Z$_{2}$
topological classification following Kane and Mele.\cite{Kane-05prlb}. In
the representation for the Dirac matrices in Eq. (\ref{Dirac-matrices}), the
time-reversal operator here is defined as\cite{Bjorken} $\Theta \equiv
-i\alpha _{x}\alpha _{z}\mathcal{K}$, where $\mathcal{K}$ the complex
conjugate operator that forms the complex conjugate of any coefficient that
multiplies a ket or wave function (and stands on the right of $\mathcal{K}$%
). Under the time reversal operation, the modified Dirac equation remains
invariant, $\Theta H(p)\Theta ^{-1}=H(-p)$ ($p$ is a good quantum number of
the momentum). Furthermore we have the relations that $\Theta
u_{1}(p)=-iu_{2}(-p)$ and $\Theta u_{2}(p)=+iu_{1}(-p)$, which satisfy the
relation of $\Theta ^{2}=-1$. Similarly, $\Theta u_{3}(p)=-iu_{4}(-p)$ and $%
\Theta u_{4}(p)=+iu_{3}(-p)$. Thus the solutions of \{$u_{1}(p)$, $u_{2}(-p)$%
\} and \{$u_{3}(p)$, $u_{4}(-p)$\} are two degenerate Kramer pairs of
positive and negative energies, respectively. The matrix of overlap $\left\{
\left\langle u_{\mu }(p)\right\vert \Theta \left\vert u_{\nu
}(p)\right\rangle \right\} $ has the form%
\begin{equation}
\left( 
\begin{array}{cccc}
0 & i\frac{mv^{2}-Bp^{2}}{E_{p,1}} & -i\frac{p_{-}v}{E_{p,1}} & i\frac{p_{z}v%
}{E_{p,1}} \\ 
-i\frac{mv^{2}-Bp^{2}}{E_{p,1}} & 0 & i\frac{p_{z}v}{E_{p,1}} & i\frac{p_{+}v%
}{E_{p,1}} \\ 
i\frac{p_{-}v}{E_{p,1}} & -i\frac{p_{z}v}{E_{p,1}} & 0 & i\frac{mv^{2}-Bp^{2}%
}{E_{p,1}} \\ 
-i\frac{p_{z}v}{E_{p,1}} & -i\frac{p_{+}v}{E_{p,1}} & -i\frac{mv^{2}-Bp^{2}}{%
E_{p,1}} & 0%
\end{array}%
\right) .
\end{equation}%
which is antisymmetric, $\left\langle u_{\mu }(p)\right\vert \Theta
\left\vert u_{\nu }(p)\right\rangle =-\left\langle u_{\nu }(p)\right\vert
\Theta \left\vert u_{\mu }(p)\right\rangle $. For the two negative energy
bands $u_{3}(p)$ and $u_{4}(p)$, the submatrix of overlap can be expressed
in terms of a single number as $\epsilon _{\mu \nu }P(p),$ 
\begin{equation}
P(\mathbf{p})=i\frac{mv^{2}-Bp^{2}}{\sqrt{(mv^{2}-Bp^{2})^{2}+v^{2}p^{2}}}.
\label{PknoV}
\end{equation}%
which is the Pfaffian for the $2\times 2$ matrix. According to Kane and Mele,%
\cite{Kane-05prlb} the even or odd number of the zeros in $P(\mathbf{p})$
defines the Z$_{2}$ topological invariant. Here we want to emphasize that
the sign of a dimensionless parameter $mB$ will determine the Z$_{2}$
invariant of the modified Dirac equation. Since $P(\mathbf{p})$ is always
non-zero for $mB\leq 0$ and there exists no zero in the Pfaffian, we
conclude immediately that the modified Dirac Hamiltonian for $mB\leq 0$
including the conventional Dirac Hamiltonian ($B=0$) is topologically
trivial.

For $mB>0$ the case is different. In this continuous model, the Brillouin
zone becomes infinite. At $p=0$ and $p=+\infty ,$ $P(0)=i$sgn$(m)$ and $%
P(+\infty )=-i$sgn$(B)$. In this case $P(\mathbf{p})=0$ at $p^{2}=mv^{2}/B$. 
$\mathbf{p}=0$ is always one of the time reversal invariant momenta (TRIM).
As a result of an isotropic model in the momentum space, we can think all
points of $p=+\infty $ shrink into one point if we regard the continuous
model as a limit of the lattice model by taking the lattice space $%
a\rightarrow 0$ and the reciprocal lattice vector $G=2\pi /a\rightarrow
+\infty $. In this sense as a limit of a square lattice other three TRIM
have $P(0,G/2)=P(G/2,0)=P(G/2,G/2)=P(+\infty )$ which has an opposite sign
of $P(0)$ if $mB>0.$ Similarly for a cubic lattice $P(\mathbf{p})$ of other
seven TRIM have opposite sign of $P(0)$. Following Fu, Kane and Mele\cite%
{Fu-07prb,Fu-07prl}, we conclude that \textit{the modified Dirac Hamiltonian
is topologically non-trivial only if }$mB>0$.

In two dimension Z$_{2}$ index can be determined by evaluating the winding
number of the phase of $P(p)$ around a loop of enclosing the half the
Brilouin zone in the complex plane of $\mathbf{p}=p_{x}+ip_{y}$, 
\begin{equation}
I=\frac{1}{2\pi i}\oint_{C}d\mathbf{p}\cdot \nabla _{\mathbf{p}}\mathrm{log}%
[P(\mathbf{p})+i\delta ].
\end{equation}%
Because the model is isotropic, the integral then reduces to only the path
along $p_{x}$ -axis while the part of the half-circle integral vanishes for $%
\delta >0$ and $\left\vert \mathbf{p}\right\vert \rightarrow +\infty $.
Along the $p_{x}$ axis one one of a pair of zeros in the ring is enclosed in
the contour C when $mB>0$, which give a Z$_{2}$ index $I=1$. This defines
the non-trivial QSH phase.

\section{Topological invariants and quantum phase transition}

An alternative approach to explore the topological property of the Dirac
model is the Green function method.\cite{Volovik-03book} Volovik \cite%
{Volovik-10xxx}proposed that the Green function rather than the Hamiltonian
is more applicable to classify the topological insulator. From the Dirac
equation, the Green function has the form%
\begin{eqnarray*}
G(i\omega _{n},p) &=&\frac{1}{i\omega _{n}-H} \\
&=&\frac{v\mathbf{p}\cdot \alpha +(mv^{2}-Bp^{2})\beta -i\omega _{n}}{\omega
_{n}^{2}+h^{2}(p)}
\end{eqnarray*}%
where $h^{2}(k)=H^{2}=v^{2}p^{2}+(mv^{2}-Bp^{2})^{2}$. There is the
following topological invariant 
\begin{equation*}
\tilde{N}=\frac{1}{24\pi ^{2}}\epsilon _{ijk}\text{Tr}[K_{i\omega _{n}=0}d%
\mathbf{p}G\partial _{p_{i}}G^{-1}G\partial _{p_{j}}G^{-1}G\partial
_{p_{k}}G^{-1}] 
\end{equation*}%
where $K=\sigma _{y}\otimes \sigma _{0}$ is the symmetry-related operator.
After tedious algebra, it is found that%
\begin{equation*}
\tilde{N}=\text{sgn}(m)+\text{sgn}(B). 
\end{equation*}%
When $mB>0$, $\tilde{N}=\pm 2$, which define the phase topologically
non-trivial. If $B$ is fixed to be positive, there exist a quantum phase
transition from topologically trivial phase of $m<0$ to a topologically
non-trivial phase. This is in a good agreement with the result of Z2 index
in the preceding section.

Except for the phases of $\tilde{N}=\pm 2$, it is found that there exist a
marginal topological phases of $\tilde{N}=\pm 1$. For free Dirac fermions of 
$B=0$, the topological invariant $\tilde{N}=sgn(m)$. It is +1 for a positive
mass and -1 for a negative mass. Their difference $\Delta \tilde{N}=2$ which
is the origin of the existence of the bound states at the interface of two
systems with positive and negative mass as we discussed in Section II. There
exists an intermediate gapless phases of $m=0$ between two topological
nontrivial ($\tilde{N}=\pm 2$) and trivial ($\tilde{N}=0$) phases. At the
critical point of topological quantum phase transition, all intermediate
states are gapless. Its topological invariant is also $\tilde{N}=+1$ or $-1$
just like as the free Dirac fermions.

\section{The topologically protected boundary states solutions}

\subsection{1D: the bound state of zero energy}

Let us start with the 1D case. In this case, the equation in Eq. (\ref%
{Dirac-eq}) can be decoupled into two sets of independent equations in the
form%
\begin{equation}
h(x)=vp_{x}\sigma _{x}+\left( mv^{2}-Bp_{x}^{2}\right) \sigma _{z}.
\end{equation}%
For a semi-infinite chain, we consider an open boundary condition at $x=0$.
We may have a series of extended solutions which spread in the whole space.
In this section we focus on the solutions of bound states near the end of
the chain. We require that the wave function vanishes at $x=+\infty $. In
the condition of $mB>0$, there exists a solution of the bound state with
zero energy%
\begin{equation}
\left( 
\begin{array}{c}
\varphi \\ 
\chi%
\end{array}%
\right) =\frac{C}{\sqrt{2}}\left( 
\begin{array}{c}
\text{sgn}(B) \\ 
i%
\end{array}%
\right) (e^{-x/\xi _{+}}-e^{-x/\xi _{-}})
\end{equation}

\begin{equation}
\xi _{\pm }^{-1}=\frac{v}{2\left\vert B\right\vert \hbar }\left( 1\pm \sqrt{%
1-4mB}\right)
\end{equation}%
where $C$ is the normalization constant. The main feature of this solution
is that the wave function dominantly distributes near the boundary. The two
parameters $\xi _{-}>\xi _{+}$ and decides the spatial distribution of the
wave function. This is a very important length scale, which characterizes
the bound state. When $B\rightarrow 0$, $\xi _{+}\rightarrow \left\vert
B\right\vert \hbar /v$ and $\xi _{-}=\hbar /mv$ i.e., $\xi _{+}$ approaches
to zero, and $\xi _{-}$ becomes a finite constant. If we relax the
constraint of the vanishing wave function at the boundary, the solution
exists even if $B=0$. In this way, we go back the conventional Dirac
equation. In this sense, the two equations reach at the same conclusion.

In the four-component form to Eq.(\ref{Dirac-eq}), two degenerate solutions
have the form, 
\begin{subequations}
\begin{eqnarray}
\Psi _{1} &=&\frac{C}{\sqrt{2}}\left( 
\begin{array}{c}
\text{sgn}(B) \\ 
0 \\ 
0 \\ 
i%
\end{array}%
\right) (e^{-x/\xi _{+}}-e^{-x/\xi _{-}}) \\
\Psi _{2} &=&\frac{C}{\sqrt{2}}\left( 
\begin{array}{c}
0 \\ 
\text{sgn}(B) \\ 
i \\ 
0%
\end{array}%
\right) (e^{-x/\xi _{+}}-e^{-x/\xi _{-}})
\end{eqnarray}

\subsection{2D: the helical edge states}

In two dimension, the equation is decoupled into two independent equations 
\end{subequations}
\begin{equation}
h_{\pm }=vp_{x}\sigma _{x}\pm vp_{y}\sigma _{y}+\left( mv^{2}-Bp^{2}\right)
\sigma _{z}.  \label{2D-Dirac}
\end{equation}%
These two subsets of equations breaks the "time" reversal symmetry under the
transformation of $\sigma _{i}\rightarrow -\sigma _{i}$ and $%
p_{i}\rightarrow -p_{i}$.

We consider a semi-infinite plane with the boundary at $x=0$. $p_{y}$ is a
good quantum number. At $p_{y}=0$, the 2D equation has the same form as the
1D equation. The x-dependent part of the solutions of bound states has the
identical form as in 1D. Thus we use the two 1D solutions \{$\Psi _{1},\Psi
_{2}$\} as the basis. The y-dependent part $\Delta H_{2D}=vp_{y}\alpha
_{y}-Bp_{y}^{2}\beta $ is regarded as the perturbation to the 1D
Hamiltonian. In this way, we have a 1D effective model for the helical edge
states%
\begin{equation}
H_{eff}=(\left\langle \Psi _{1}\right\vert ,\left\langle \Psi
_{2}\right\vert )\Delta H\left( 
\begin{array}{c}
\left\vert \Psi _{1}\right\rangle \\ 
\left\vert \Psi _{2}\right\rangle%
\end{array}%
\right) =vp_{y}\text{sgn}(B)\sigma _{z}
\end{equation}%
The sign dependence of $B$ in the effective model also reflects the fact
that the helical edge states disappear if $B=0$. The dispersion relations
for the bound states at the boundary are%
\begin{equation}
\epsilon _{p}=\pm vp_{y}
\end{equation}%
Electrons will have positive ($+v$) and negative velocity ($-v$) in two
different states, respectively, and form form a pair of helical edge states.
Thus the 2D equation can describe a quantum spin Hall system.

The exact solutions of the edge states to this 2D equation have the similar
form of 1D\cite{Zhou-08prl} 
\begin{subequations}
\begin{eqnarray}
\Psi _{1} &=&\frac{C}{\sqrt{2}}\left( 
\begin{array}{c}
\text{sgn}(B) \\ 
0 \\ 
0 \\ 
i%
\end{array}%
\right) (e^{-x/\xi _{+}}-e^{-x/\xi _{-}})e^{+ip_{y}y/\hbar } \\
\Psi _{2} &=&\frac{C}{\sqrt{2}}\left( 
\begin{array}{c}
0 \\ 
\text{sgn}(B) \\ 
i \\ 
0%
\end{array}%
\right) (e^{-x/\xi _{+}}-e^{-x/\xi _{-}})e^{+ip_{y}y/\hbar }
\end{eqnarray}%
with the dispersion relation $E_{p_{y}}=\pm vp_{y}$sgn$(B)\sigma _{z}$. The
penetration depth becomes $p_{y}$ dependent, 
\end{subequations}
\begin{equation}
\xi _{\pm }^{-1}=\frac{v}{2\left\vert B\right\vert \hbar }\left( 1\pm \sqrt{%
1-4mB+4B^{2}p_{y}^{2}/v^{2}}\right) .
\end{equation}

In two-dimension, the Chern number or Thouless-Kohmoto-Nightingale-Nijs
integer can be used to characterize whether the system is topologically
trivial or non-trivial.\cite{TNNK} Write the Hamiltonian in Eq. (\ref%
{2D-Dirac}) in the form $H=\mathbf{d}(p)\cdot \sigma $ The Chern number is
expressed as%
\begin{equation*}
n_{c}=\int \frac{d\mathbf{p}}{(2\pi \hbar )^{2}}\frac{\epsilon _{ijk}d_{i}%
\frac{\partial d_{j}}{\partial p_{x}}\frac{\partial d_{k}}{\partial p_{y}}}{%
d^{3}} 
\end{equation*}%
where $d^{2}=\sum_{\alpha =x,y,z}d_{\alpha }^{2}$.\cite{TNNK,Zhou-06prb} The
integral runs over the first Brillouin zone for a lattice system. The number
is always an integer for an finite first Brillouin zone, but can be
fractional for an infinite zone. For these two equations the Chern number
has the form \cite{Lu-10prb,Shan-10njp}%
\begin{equation}
n_{\pm }=\pm (\text{sgn}(m)+\text{sgn}(B))/2.  \label{TKNN}
\end{equation}%
which gives the Hall conductance $\sigma _{\pm }=n_{\pm }e^{2}/h$. When $m$
and $B$ have the same sign, $n_{\pm }$ becomes $\pm 1$, and the systems are
topologically non-trivial. But if $m$ and $B$ have different signs, $n_{\pm
}=0$. The topologically non-trivial condition is in agreement with the
existence condition of edge state solution. This reflects the bulk-edge
relation of integer quantum Hall effect.\cite{Hatsugai-93prl}

\subsection{3D: the surface states}

In 3D, we consider an $y$-$z$ plane at $x=0$. We can derive an effective
model for the surface states by means of the 1D solutions of the bound
states. Consider $p_{y}$- and $p_{z}$-dependent part as a perturbation to 1D 
$H_{1D}(x)$, 
\begin{equation}
\Delta H_{3D}=vp_{y}\alpha _{y}+vp_{z}\alpha
_{z}-B(p_{y}^{2}+p_{z}^{2})\beta .
\end{equation}%
The solutions of 3D Dirac equation at $p_{y}=p_{z}=0$ are identical to the
two 1D solutions. A straightforward calculation as in the 2D case gives%
\begin{equation}
H_{eff}=(\left\langle \Psi _{1}\right\vert ,\left\langle \Psi
_{2}\right\vert )\Delta H_{3D}\left( 
\begin{array}{c}
\left\vert \Psi _{1}\right\rangle \\ 
\left\vert \Psi _{2}\right\rangle%
\end{array}%
\right) =v\text{sgn}(B)(p\times \sigma )_{x}.
\end{equation}%
Under a unitary transformation, 
\begin{subequations}
\begin{eqnarray}
\Phi _{1} &=&\frac{1}{\sqrt{2}}(\left\vert \Psi _{1}\right\rangle
-i\left\vert \Psi _{2}\right\rangle ) \\
\Phi _{2} &=&\frac{-i}{\sqrt{2}}(\left\vert \Psi _{1}\right\rangle
+i\left\vert \Psi _{2}\right\rangle )
\end{eqnarray}%
we can have a gapless Dirac equation for the surface states

\end{subequations}
\begin{eqnarray}
H_{eff} &=&\frac{1}{2}(\left\langle \Phi _{1}\right\vert ,\left\langle \Phi
_{2}\right\vert )\Delta H_{3D}\left( 
\begin{array}{c}
\left\vert \Phi _{1}\right\rangle \\ 
\left\vert \Phi _{2}\right\rangle%
\end{array}%
\right)  \notag \\
&=&v\text{sgn}(B)(p_{y}\sigma _{y}+p_{z}\sigma _{z}).
\end{eqnarray}%
The dispersion relations become $E_{p}=\pm vp.$ In this way we have an
effective model for a single Dirac cone of the surface states.

The exact solutions of the surface states to this 3D equation with the
boundary are 
\begin{subequations}
\begin{equation}
\Psi _{\pm }=C\Psi _{\pm }^{0}(e^{-x/\xi _{+}}-e^{-x/\xi _{-}})\exp
[+i\left( p_{y}y+p_{z}z\right) /\hbar ]
\end{equation}%
where 
\begin{eqnarray}
\Psi _{+}^{0} &=&\left( 
\begin{array}{c}
\cos \frac{\theta }{2}\text{sgn}(B) \\ 
-i\sin \frac{\theta }{2}\text{sgn}(B) \\ 
\sin \frac{\theta }{2} \\ 
i\cos \frac{\theta }{2}%
\end{array}%
\right) \\
\Psi _{-}^{0} &=&\left( 
\begin{array}{c}
\sin \frac{\theta }{2}\text{sgn}(B) \\ 
i\cos \frac{\theta }{2}\text{sgn}(B) \\ 
-\cos \frac{\theta }{2} \\ 
i\sin \frac{\theta }{2}%
\end{array}%
\right)
\end{eqnarray}%
with the dispersion relation $E_{\pm }=\pm vp$sgn$(B)$ and $p=\sqrt{%
p_{y}^{2}+p_{z}^{2}}$. The penetration depth becomes $p$ dependent, 
\end{subequations}
\begin{equation}
\xi _{\pm }^{-1}=\frac{v}{2\left\vert B\right\vert \hbar }\left( 1\pm \sqrt{%
1-4mB+4B^{2}p^{2}/\hbar ^{2}}\right) .
\end{equation}

\subsection{Generalization to higher dimensional topological insulators}

The solution can be generalized to higher dimensional system. We conclude
that there always exists a d-dimensional surface state in the modified Dirac
equation.

\section{Application to real systems}

Now we address the relevance of the modified Dirac model to real materials.
Of course we cannot simply apply the Dirac equation to semiconductors
explicitly. Usually the band structures of most semiconductors or others
have no particle-hole symmetry. Thus a quadratic term should be introduced
into the modified Dirac model. On the other hand the band structure may not
be isotropic and the effective velocities along different axes are
different. A more general model has the form,%
\begin{equation}
H=\epsilon (p_{i})+\sum_{i}v_{i}p_{i}\alpha
_{i}+(mv^{2}-\sum_{i}B_{i}p_{i}^{2})\beta .  \label{general-Dirac}
\end{equation}%
To have a solution for topological insulator, the additional terms must keep
the band gap open. Otherwise it cannot describe an insulator.

The equation in solids can be derived from the theory of invariant or the k$%
\cdot $p theory as an expansion of the momentum p near the $\Gamma $ point.
Since under the time reversal, $\beta \rightarrow \beta $ and $\alpha
\rightarrow -\alpha $, if we expand an time reversal invariant Hamiltonian
near the $\Gamma $ point, the zero-order term should be constant, $\epsilon
(0)$ and $mv^{2}\beta $. The first order term in the momentum must $%
\sum_{i}v_{i}p_{i}\alpha _{i}$ since $p_{i}\rightarrow -p_{i}$ under time
reversal. The second order term is $\sum_{i}B_{i}p_{i}^{2}\beta $ and $\frac{%
p^{2}}{2m}$ in $\epsilon (p_{i})$. The third order term is the cubic term in 
$\alpha $. The summation up to the second order terms give the modified
Dirac equation.

\subsection{Complex p-wave spinless superconductor}

A complex p-wave spinless superconductor has two topologically distinct
phases, one is the strong pairing phase and another is the weak pairing
phase.\cite{Read-00prb,Volovik-03book} The weak pairing phase is identical
to the Moore-Read quantum Hall state.\cite{Read-00prb} The system can be
described by the modified Dirac model. In the BCS mean field theory, the
effective Hamiltonian for quasiparticles in this system has the form%
\begin{equation}
K_{eff}=\sum_{k}\left[ \xi _{k}c_{k}^{\dag }c_{k}+\frac{1}{2}(\Delta
_{k}^{\ast }c_{-k}c_{k}+\Delta _{k}c_{k}^{\dag }c_{-k}^{\dag })\right] .
\end{equation}%
The normalized ground state has the form%
\begin{equation}
\left\vert \Omega \right\rangle =\prod_{k}(u_{k}+v_{k}c_{k}^{\dag
}c_{-k}^{\dag })\left\vert 0\right\rangle .
\end{equation}%
where $\left\vert 0\right\rangle $ is the vacuum state. The Bogoliubov-de
Gennes equation for $u_{k}$ and $v_{k}$ becomes%
\begin{equation}
i\hbar \frac{\partial }{\partial t}\left( 
\begin{array}{c}
u_{k} \\ 
v_{k}%
\end{array}%
\right) =\left( 
\begin{array}{cc}
\xi _{k} & -\Delta _{k}^{\cdot } \\ 
-\Delta _{k} & -\xi _{k}%
\end{array}%
\right) \left( 
\begin{array}{c}
u_{k} \\ 
v_{k}%
\end{array}%
\right)
\end{equation}%
For complex p-wave pairing, we take $\Delta _{k}$ to be an eigen function of
rotations in $k$ of eigenvalue of two-dimensional angular momentum! $l=-1$,
and thus at small $k$ it generically takes the form%
\begin{equation}
\Delta _{k}=\Delta (k_{x}-ik_{y});\xi _{k}=\frac{k^{2}}{2m}-\mu
\end{equation}%
In this way the Bogoliubov-de Gennes equation has the exact form of 2D
modified Dirac equation%
\begin{equation}
H_{eff}=-\Delta \left( k_{x}\sigma _{x}+k_{y}\sigma _{y}\right) +\left( 
\frac{k^{2}}{2m}-\mu \right) \sigma _{z}
\end{equation}%
The Chern number of the effective Hamiltonian becomes%
\begin{equation}
n=\left[ \text{sgn}(\mu )+\text{sgn}(1/m)\right] /2
\end{equation}%
Since we assume the mass of the spinless particles $m$ positive, we conclude
that for a positive $\mu (>0)$ the Chern number is $+1$ and for a negative $%
\mu $ the Chern number is $0$. For $\mu =0$, the Chern number is equal to
one half, which is similar to the case of $m\rightarrow +\infty $ and a
finite $\mu .$ If the quadratic term in $\xi _{k}$ is neglected, we see that
the topological property will change completely.

Usually for a positive $\mu $, the system is in a weak pairing phase, for a
negative $\mu $ the strong coupling phase. Including the quadratic term in $%
\xi _{k}$ we conclude that the weak pairing phase for positive $\mu $ is a
typical topological insulator, which possesses a chiral edge state if the
system has a boundary. The exact solution of this equation can be found in
the paper by Zhou et al.\cite{Zhou-08prl} Read and Green\cite{Read-00prb}
argued that a bound state solution exists at a straight domain wall parallel
to the y-axis, with $\mu (r)=\mu (x)$ small and positive\ for $x>0$, and
negative for $x<0$. There is only one solution for each $k_{y}$ and so we
have a chiral Majorana fermions on the domain wall. From the 2D solution,
the system in a weak pairing phase should have a topologically protected and
chiral edge state of Majorana fermion. Recently Fu and Kane proposed that as
a superconducting proximity effect the interface of the surface state of
three-dimensional topological insulator and an s-wave superconductor
resembles a spinless $p_{x}+ip_{y}$ superconductor, but does not break time
reversal symmetry.\cite{Fu-08prl} The state support Majorana bound states at
vortices.

\subsection{Quantum Spin Hall Effect: HgTe/CdTe quantum well and thin film
of topological insulator}

In 1988 Haldane proposed a spinless fermion model for IQHE without Landau
levels, in which two independent effective Hamiltonian with the same form of
2D the Dirac equation were obtained.\cite{Haldane-88prl} The Haldane's model
was generalized to the graphene lattice model of spin 1/2 electrons, which
exhibits quantum spin Hall effect.\cite{Kane-05prl} Bernevig, Hughes and
Zhang predicted that QSHE can be realized in HgTe/CdTe quantum well and
proposed an effective model,\cite{Bernevig-06science} 
\begin{equation}
H_{BHZ}=\left( 
\begin{array}{cc}
h(k) & 0 \\ 
0 & h^{\ast }(-k)%
\end{array}%
\right)
\end{equation}%
where $h(k)=\epsilon (k)+A(k_{x}\sigma _{x}+k_{y}\sigma
_{y})+(M-Bk^{2})\sigma _{z}$. The model is actually equivalent to the 2D
Dirac model as shown in Eq.(\ref{2D-Dirac}) in addition of the kinetic term $%
\epsilon (k)$, 
\begin{equation}
h(k)=\epsilon (k)+h_{+};h^{\ast }(-k)=\epsilon (k)+Uh_{-}U^{-1},
\end{equation}%
where the unitary transformation matrix $U=\sigma _{z}$.

If the inclusion of $\epsilon (k)$ does not close the energy gap caused by $%
M $ for a non-zero $B$, there exists a topological phase transition from a
positive $M$ to a negative $M$. However, the sign of $M$ alone cannot
determine whether the system is topologically trivial or non-trivial. From
the formula in Eq.(\ref{TKNN}), we know that the system is in a quantum spin
Hall phase only for $MB>0$ and there exists a pair of helical edge states
around the boundary of system. A general discussion can be found in the
paper by Zhou et al.\cite{Zhou-08prl} Finally we want to comment on one
popular opinion that the band inversion induces the topological quantum
phase transition. If $B=0$, the system is always topologically trivial for
either positive or negative $M$, though there exists a bound state at the
interface of two systems with positive and negative $M$, respectively.

The surface states of a thin film of topological insulator such as Bi$_{2}$Te%
$_{3}$ and Bi$_{2}$Se$_{3}$ can be also described by a two-dimensional Dirac
model.\cite{Lu-10prb,Shan-10njp} The mass or the band gap of the Dirac
particles originates from the overlapping of the wave functions of the top
and bottom surface states. The gap opening of the two surface states were
observed in Bi$_{2}$Se$_{3}$ thin films experimentally\cite%
{Zhang-10np,Sakamoto-10prb}, and was also confirmed numerically by DFT\cite%
{Park-10prl}. Recently Luo and Zunger \cite{Luo-10prl} reported a DFT
calculation for HgTe/CdTe quantum well and presented a different picture
that the topological quantum phase transition occurs at the crossing point
of two "interface-localized" states. This is in a good agreement of the
theory for 3D topological insulator thin film.\cite{Chu-10xxx}

\subsection{Three-Dimensional Topological Insulators}

The 3D Dirac equation can be applied to describe a large family of
three-dimensional topological insulators. Bi$_{2}$Te$_{3}$ and Bi$_{2}$Se$%
_{3}$ and Sb$_{2}$Te$_{3}$ have been confirmed to be topological insulator
with a single Dirac come of surface states. For example, in Bi$_{2}$Te$_{3}$%
, the electrons near the Fermi surfaces mainly come from the p-orbitals of
Bi and Te atoms. According to the point group symmetry of the crystal
lattice, $p_{z}$ orbital splits from $p_{x,y}$ orbital. Near the Fermi
surface the energy levels turn out to be the $p_{z}$ orbital. The four
orbitals are used to construct the eigenstates of parity and the base for
the effective Hamiltonian,\cite{Zhang-09NP} which has the exact form as 
\begin{equation}
H=\epsilon (k)+\sum_{i=x,y,z}v_{i}p_{i}\alpha
_{i}+(mv^{2}-\sum_{i=x,y,z}B_{i}p_{i}^{2})\beta .
\end{equation}%
with $v_{x}=v_{y}=v_{\shortparallel }$ and $v_{z}=v_{\perp }$ and $%
B_{x}=B_{y}=B_{\shortparallel }$ and $B_{z}=B_{\perp }$. $\epsilon
(k)=C-D_{\shortparallel }(p_{x}^{2}+p_{y}^{2})-D_{\perp }p_{z}^{2}$. In this
way the effective Hamiltonian in the x-y plane has the form\cite{Shan-10njp}

\begin{equation}
H_{eff}=\sqrt{1-D_{\perp }^{2}/B_{\perp }^{2}}v_{\shortparallel }(p\times
\sigma )_{z}
\end{equation}%
or under a unitary transformation

\begin{equation}
H_{eff}=\sqrt{1-D_{\perp }^{2}/B_{\perp }^{2}}v_{\shortparallel
}(p_{x}\sigma _{x}+p_{y}\sigma _{y}).
\end{equation}%
We note that the inclusion of $\epsilon (k)$ will revise the effective
velocity of the surface states, which is different from the result in Ref.%
\cite{Zhang-09NP}.

\section{From the continuous model to the lattice model}

In practice, the continuous model is sometimes mapped into a lattice model
in the tight binding approximation. In a d-dimensional hyper-cubic lattice,
one replaces\cite{Li-05prb,Qi-06prb} 
\begin{eqnarray}
k_{i} &\rightarrow &\frac{1}{a}\sin k_{i}a \\
k_{i}^{2} &\rightarrow &\frac{2}{a^{2}}(1-\cos k_{i}a)
\end{eqnarray}%
which are equal to each other in a long wave limit. Usually there exits the
fermion doubling problem in the lattice model for massless Dirac particles.
The replacement of $k_{i}\rightarrow \sin k_{i}a/a$ will cause an additional
zero point at $k_{i}a=\pi $ besides $k_{i}a=0$. Thus there exist two Dirac
cones in a square lattice at $k=(0,0)$ and ($\pi /a,\pi /a$) for a gapless
Dirac equation. A large $B$ term removes the problem as $2B(1-\cos
k_{i}a)/a^{2}\rightarrow B/a^{2}$ in the lattice model. Thus the lattice
model is equivalent to the continuous model only in the condition of a large 
$B$. The zero point of $(1-\cos k_{i}a)^{2}$ is at $k_{i}a=\pi /2$ not 0 or $%
\pi $ in $\sin k_{i}a$. Thus for a finite $B$, the band gap may not open at
the $\Gamma $ point in the lattice model because of the competition between
the linear term and the quadratic term of $k_{i}$. This fact may lead to a
topological transition from a large $B$ to a small $B$. Imura et al \cite%
{Imura-10prb} analyzed the 2D case in details and found that there exists a
topological transition at a finite value of $B$ in two-dimension. A similar
transition will also exist in higher dimension. It should be careful when we
study the continuous model in a tight binding approximation.

\section{Conclusion}

To summarize, we found that the Z$_{2}$ index for the Dirac equation is
always zero, and thus the Dirac equation is topologically trivial. After the
quadratic $B$ term is introduced to correct the mass $m$ of the Dirac
equation, the Z$_{2}$ index is modified as 1 for $mB>0$ and 0 for $mB<0$.
For a fixed $B$ there exists a topological quantum phase transition from a
topologically trivial system to a non-trivial one system when the sign of
mass $m$ changes.

From the solutions of the modified Dirac equation, we found that under the
condition of $mB>0$,

\begin{itemize}
\item in 1D, there exists the bound state of zero energy near the boundary;

\item in 2D, there exists the solution of helical edge states near the
boundary;

\item in 3D, there exists the solution of the surface states near the
surface;

\item in higher dimension, there always exists the solution of higher
dimension surface.
\end{itemize}

From the solutions of the bound states near the boundary, and the
calculation of Z$_{2}$ index we conclude that the modified Dirac equation
can provide a description of a large families of topological insulators from
one to higher dimension.\newline

\textbf{Acknowledgements}

This work was supported by the Research Grant Council of Hong Kong under
Grant No. HKU7051/10P and HKUST3/CRF/09.

\end{document}